\documentclass[conference]{IEEEtran}

\usepackage{graphicx}
\usepackage{amsfonts}
\usepackage[cmex10]{amsmath}
\usepackage{cite}
\usepackage{algorithmic}
\usepackage[center]{caption}
\usepackage{epsfig}
\usepackage{latexsym}
\usepackage{epstopdf}
\usepackage{verbatim}
\usepackage{amsthm}
\usepackage[linesnumbered,ruled,vlined]{algorithm2e}
 
\graphicspath{{./figures/}}

\begin{document}
\title{A Pricing-Based Cooperative Spectrum Sharing Stackelberg Game}
\author{\large Ramy E. Ali$^\dagger$, Karim G. Seddik$^\ddagger$, Mohammed Nafie$^\dagger$, and Fadel F. Digham$^\star$ \\ [.1in] 
\small  
\begin{tabular}{c} $^\dagger$Wireless Intelligent
Networks Center (WINC), Nile University, Smart Village, Egypt.\\
$^\ddagger$Electronics Engineering Department, American University in Cairo, AUC Avenue, New Cairo 11835, Egypt.\\
$^\star$National Telecom Regulatory Authority (NTRA), Egypt \\
email: ramy.essam@nileu.edu.eg, kseddik@aucegypt.edu, mnafie@nileuniversity.edu.eg, fadel.digham@ieee.org
\end{tabular} 
}

\maketitle

\begin{abstract}
In this paper, we study the problem of cooperative spectrum
sharing among a primary user (PU) and multiple secondary users (SUs) under quality of service (QoS) constraints. The SUs network is controlled by the PU through a relay which gets a revenue for amplifying and forwarding the SUs' signals to their respective destinations. The relay charges each SU a different price depending on its received signal-to-interference-and-noise ratio (SINR). The primary relay controls the SUs network and maximize any desired PU utility function. The PU utility function represents its QoS, which is affected by the SUs access, and its gained revenue to allow the access of the SUs. The problem of maximizing the primary utility is formulated as a Stackelberg game and solved through three different approaches, namely, the optimal, the heuristic and the suboptimal algorithms.
\end{abstract}


\IEEEpeerreviewmaketitle
\section{Introduction}
\makeatletter{\renewcommand*{\@makefnmark}{}
\footnotetext{\hrule \vspace{0.05in} This work was supported by a grant from the Egyptian National Telecommunications Regulatory Authority (NTRA).

Mohamed Nafie is also affiliated with the EECE
Dept., Faculty of Engineering, Cairo University.
}\makeatother }

Cognitive radio (CR) is a promising technology which can
enhance the spectrum utilization efficiency by allowing the
secondary usage of the under-utilized licensed spectrum held by primary users (PUs)\cite{akyildiz2006next}, \cite{haykin2005cognitive}. To utilize the spectrum holes, cooperative spectrum sharing allows the secondary users (SUs) to make use of the PU licensed spectrum as long as their interference to the PUs does not exceed a predefined threshold set by the PU. In return, the PUs would earn some money or use the SU as a cooperative relay to improve their transmission, so a win-win situation can be achieved.

Game theory is a powerful tool which can be used to study and analyze the competition between the users willing to access the spectrum \cite{osborne1994course}, \cite{felegyhazi2006game}. In \cite{wang2010cooperative}, a cooperative spectrum sharing approach was proposed in which the PU selects a set of SUs as the cooperative relays for its transmission. In return, the PU leases portion of channel access time to the selected SUs for their own transmission. The access time of each SU is proportional to its contribution in the PU transmission. The SUs game is investigated as a non-cooperative game.

In\cite{hao2011stackelberg}, a Stackelberg game \cite{osborne1994course}  was considered, in which the PU plays the role of the leader and SUs are the followers. The primary transmitter (PT) may select a secondary transmitter (ST) as a cooperative relay or not depending on the PT desired rate. The PU allows the access of the SUs to its spectrum part of the time in a random access manner. A ST should make a payment to the PU depending on the probability with which it attempts to access the channel. The ST which is selected as a cooperative relay pays less than the other SUs. Thus a win-win situation can be achieved.  In \cite{wang2012relay}, a cognitive radio network of one PU, a relay and one SU was considered. A relay assisted spectrum sharing scheme based on the mixed sharing strategy was proposed, in which the ST adapts its power according to the sensing results of the PU spectrum. If the PT is sensed to be OFF, the ST transmits with a higher power which maximizes its rate. If the PU is sensed to be ON, the ST transmits with a power below the interference threshold of the PU to the relay then the relay decodes and forwards the the ST signal to the SD.

 In \cite{singh2012interference}, a cognitive radio network with multiple SUs and one PU is considered. The SUs power control problem is formulated as a sum-rate maximization problem under PU and SU quality of service (QoS). A convex approximation approach is introduced through an iterative algorithm which approximates this non-convex rate maximization problem as a geometric program. In this model the PU always transmits its data with a fixed power and the SUs are assumed to be non-selfish, so they transmit their data according to the power allocation vector which maximizes the overall sum-rate.

In \cite{ren2011pricing}, a traditional (non-cognitive) wireless relay network consisting of one relay node and multiple source-destination pairs was considered. Each user acts as a self-interested player, which aims at maximizing its own benefit by choosing the optimal transmit power. The competition among the users is modelled as a non-cooperative game. The relay can set prices to maximize either its revenue or any desirable system utility, and the payment of each user to the relay depends on the received signal-to-interference-and-noise ratio (SINR). In this model, the relay does not ensure a certain QoS to any of the users and the relay is mainly concerned about its revenue.

In this paper, we consider relay-assisted cognitive radio with one PU, one primary relay, and a network of $N$ selfish SUs. The transmission of the SUs is established through the primary relay which adopts the Amplify and Forward (AF) \cite{laneman2004cooperative} relaying technique. The PU adapts its transmit power, the relay power and control the SUs power allocation through the relay to maximize its utility function. Specifically, the relay adopts the differentiated pricing technique proposed in \cite{ren2011pricing} to enforce all SUs to transmit with some desired power levels that maximize the PU utility. The PU utility function is defined such that it captures the interest of the PU to maximize its QoS and the gained revenue from allowing the SUs to access its spectrum.

The main contributions of this work are summarized as follows:
\begin{itemize}
\item We formulate the cooperative spectrum sharing power control problem
as a Stackelberg game between the PU and the SUs. 
\item We propose a combinatorial optimal power control solution for the problem of maximizing the PU utility under a minimum SUs QoS requirements.
\item We also propose a simple heuristic real-time algorithm, which allows the access of a maximum of one SU.
\item Finally, we propose a low complexity suboptimal scheme which may allow more than one SU to access the spectrum.
\end{itemize}
The rest of this paper is organized as follows. In Section \ref{Model}, we present the system model. In Section \ref{Problem}, we formulate the problem of maximizing the PU utility function as a Stackelberg game. Simulation results are presented in Section \ref{Simulation}. Finally, concluding remarks are drawn in Section \ref{Discussion}.
\section{System Model} 
\label{Model}

\begin{figure}
  \centering
\includegraphics[width=.8\linewidth,height=.3\textheight]{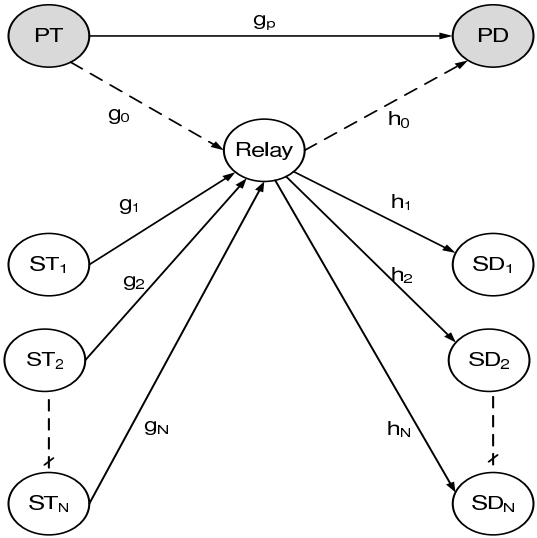} 
\caption{System Model \label{fig 1}}
\end{figure}

We consider a primary network composed of a PT, its intended destination (PD) and a relay ($\mathcal{R}$). In addition, we consider a secondary network with $N$ source-destination pairs (ST, SD). Fig.\ref{fig 1} depicts the system under consideration.  Transmission is divided into two slots (one frame). The first slot is used by all STs to transmit their signals to the relay. The $ i $-th ST, denoted by $ \text{ST}_{i} $, transmits with a power $p_{i}$, while in the second slot, the relay amplifies and forwards the received signals from all STs with a power $p_{\mathcal{R}}$ to their destinations. The PU transmits with power $p_0$ in the first slot and with power $p_{0_{max}}$ in the second slot in which its transmission is subjected to the relay interference\footnote{In the second time slot, the PU must transmit using its maximum power to maximize its rate.}.

We assume a Rayleigh flat-fading channels, which means that the channel gain of a link remains constant during one frame (two time slots). Specifically, we denote the coefficients for PT-PD channel by $g_p$, PT-$\mathcal{R}$ and the $\mathcal{R}$-PD channels by $g_0$ and $h_0$, the $\text{ST}_i$-$\mathcal{R}$ and the $\mathcal{R}$-$\text{SD}_i$ channels by $g_i$ and $h_i$, respectively. The direct links between ST$_i$-SD$_i$, ST$_i$-PD and PT-SD$_i$ are neglected due to shadowing and the too large separation \cite{ren2011pricing},\cite{rankov2007spectral}.

We assume that each SU$_i$ has a maximum power of $p_{i_{max}}$. We also assume that the relay has a variable power $p_{\mathcal{R}}$ and a maximum power of $p_{\mathcal{R}_{max}}$, unlike the assumption in \cite{ren2011pricing} which assumes that the relay always transmits with a fixed power level. We also assume that the relay has complete information about the network, i.e., channel gains and maximum power constraints. The received signal at the relay can be expressed as
\begin{equation}
y_{\mathcal{R}}=g_0\sqrt p_0x_{p_1}+ \sum_{j=1}^{N} g_j \sqrt{p_j}x_j+n_{\mathcal{R}},
\end{equation}
where $x_{p_1}$ is the unit-power transmit signal from PT to PD in the first slot, $x_j$ is the unit-power transmit signal from ST$_j$ to SD$_j$ and $n_{\mathcal{R}}$ is zero-mean additive white Gaussian noise (AWGN) with variance $N_\circ$.
The received signal at the $i$-th SU's destination ($SD_i$) can be expressed as
\begin{equation}
y_i=\alpha h_i y_{\mathcal{R}} +n_i  \  \  \ i=1,\cdots,N
\end{equation}
where $n_i$ is zero-mean AWGN with variance $N_\circ$ and $\alpha$ is the amplification factor and is given by
\begin{equation}
\alpha=\sqrt{\frac{p_{\mathcal{R}}}{\sum_{j=0}^{N} \ |g_j|^2 p_j+N_o}}.
\end{equation}
We can express the received SINR at the $i$-th SU destination, $\gamma_i(\mathbf{p})$, as 
\begin{equation}
\begin{split}
\gamma_{i}&(\mathbf{p})=\\ &\frac{|g_i|^2 |h_i|^2 p_{\mathcal{R}} p_i}{|g_i|^2 N_o p_i+\left(|h_i|^2 p_{\mathcal{R}}+N_o\right)\cdot\left(\sum_{j=0,j\neq i}^N |g_j|^2 p_j+N_o\right)},
\end{split}
\end{equation}
where $\mathbf{p}$ is the power allocation vector, which is defined as
\begin{equation*}
\mathbf{p}=\left[ p_0, p_1,\cdots, p_N, p_{\mathcal{R}}\right]^T.
\end{equation*}
The rate at which the $i$-th SU transmits is given by
\begin{equation}
R_i=\frac{1}{2} \log(1+\gamma_i) \ \ {\rm nats/sec}
\end{equation}
where the scaling factor $1/2$ is due to the fact that each SU transmits its data to the relay and remains silent in the next slot while the relay forwarding its data to the corresponding destination. We can also define the SUs's sum rate $R_{\Sigma}$ as follows
\begin{equation}
R_{\Sigma}=\sum_{i=1}^N R_i \ \ {\rm nats/sec}.
\end{equation}
The received signal at the PD in the first time slot can be expressed as
\begin{equation*}
y_{p_1}=g_p\sqrt p_0x_{p_1}+n_0,
\end{equation*}
where $n_0$ is zero-mean AWGN with variance $N_\circ$. The SINR at the PD in the first slot, $\gamma_{p_1}$, is then given by
\begin{equation}
\gamma_{p_1}=\frac{p_0|g_p|^2}{N_o}.
\end{equation}
The received signal at the PD in the second time slot can be expressed as
\begin{equation}
y_{p_2}=g_p \sqrt p_{0_{max}}x_{p_2}+\alpha h_0 y_{\mathcal{R}}+n_0,
\end{equation}
where $x_{p_2}$ is the unit-power transmit signal from PT to PD in the second slot. Similarly, the SINR at the PD in the second slot, $\gamma_{p_2}$, is given by
\begin{equation}
\gamma_{p_2}=\frac{p_{0_{max}}|g_p|^2}{N_o+p_{\mathcal{R}} |h_0|^2}.
\end{equation}
The PU rate can be averaged as follows.
\begin{equation}
\begin{split}
R_p&=\frac{1}{2} (\log(1+\gamma_{p_1})+\log(1+\gamma_{p_2}))\\
   &=\frac{1}{2} \log((1+\gamma_{p_1})(1+\gamma_{p_2}))\\ 
   &=\frac{1}{2} \log(1+\gamma_{p_e}) \ \ {\rm nats/sec},
   \end{split}
\end{equation}
where $\gamma_{p_e}$ is the PU effective SINR and is given by
\begin{equation}
\gamma_{p_e}=\gamma_{p_1}+\gamma_{p_2}+\gamma_{p_1}\gamma_{p_2}.
\end{equation}
The maximum SINR of the PU, $\gamma_{p_{max}}$, is defined as
\begin{equation}
\gamma_{p_{max}}=\frac{p_{0_{max}}|g_p|^2}{N_{\circ}}.
\end{equation}
Thus, we can also define the PU rate when all SUs are inactive, which is an upper bound for $R_p$ as
\begin{equation}
R_{p_{max}}=\log\left(1+\frac{p_{0_{max}}|g_p|^2}{N_{\circ}}\right) \ \ {\rm nats/sec}.
\end{equation}

\section{Stackelberg Game Analysis}
 \label{Problem}
The problem of  maximizing the PU cost function can be addressed as a Stackelberg game. The PU, which owns the licensed spectrum, plays the role of the leader and the SUs are the followers of this game. The PU selects the value of a weight parameter ($w_p$), $p_0$, $p_{\mathcal{R}}$ and the prices vector ($\boldsymbol{\pi}$), which contains the price that each SU$_i$ will charge to access the spectrum, then each SU$_i$ selects its transmit power $p_i$ accordingly in a non-cooperative game. Our objective is to get the Nash Equilibrium (NE) for this Stackelberg game, where neither the PU nor any of the SUs have incentive to deviate unilaterally from this NE point (Stackelberg Equilibria).

The PU may be concerned with its QoS rather than its gained revenue from the secondary network or vice versa. Hence, the PU utility function, $U_p$, can be defined as
\begin{equation}
U_p=w_p(1+\gamma_{p_1})(1+\gamma_{p_2})+R_v,
\end{equation}
where $w_p$ is a weight parameter that converts the term $(1+\gamma_{p_1})(1+\gamma_{p_2})$ into currency. The term $(1+\gamma_{p_1})(1+\gamma_{p_2})$ can be interpreted as $e^{2R_p}$ or as $(1+\gamma_{p_e})$  .The parameter $w_p$ controls the PU trade-off between its QoS and its gained revenue, and it ranges from zero, where the PU only cares about the revenue it gets from the secondary network, to infinity, where the PU only cares about its QoS. The SUs payment is a reimbursement of the PU SINR or QoS degradation caused by
the SUs.

 The term $R_v$ is the PU revenue gained from the secondary network and can be expressed as
\begin{equation}
R_v=\sum_{i=1}^N \ \pi_i\gamma_i,
\end{equation}
where $\pi_i$ is the price for SU$_i$ set by the PU.\\
The non-cooperative SUs level game, $G_{SU_s}$ is defined as 
\begin{equation}
\begin{aligned}
G_{SU_s}=\left\{ \Omega , {\left\{\mathcal{P}_i \right\}_{i \in \Omega}}, {\left\{U_{s_i} \right\}_{i \in \Omega}} \right\},
\end{aligned}
\end{equation}
where $\Omega$ is the set of all SUs and $\mathcal{P}_i$ is the allowable power strategies of the SU$_i$ which is defined as ${\mathcal{P}_i=\left\{p_i: 0 \leq p_i \leq p_{i_{max}} \right\}} $. 

The term $U_{s_i}$ is the SU$_i$ cost function which is defined as
\begin{equation}
U_{s_i}=w_sR_i-\pi_i \gamma_i,
\end{equation}
where $w_s$ is a factor that converts the rate units to currency. For simplicity, it is assumed that $w_s=1$ in the following analysis. The term $\pi_i\gamma_i$ represents the secondary payment to the PU for allowing this SU$_i$ to access the spectrum, which is a function of the received SINR, $\gamma_i$. In \cite{ren2011pricing}, it is proved that the relay can set its prices according to equation (\ref{price}), to enforce the NE \cite{felegyhazi2006game} of $G_{SU_s}$ to any desired NE, i.e, obligate all SUs to send according to any desired power allocation vector $\bar{\textbf{p}}$. 
 \begin{equation}
 \label{price}
\pi_i=\frac{1}{2(1+\gamma_i(\bar{\textbf{p}}))} \ i=1,\cdots,N,
\end{equation} 
where $\bar{\textbf{p}}=[\bar{p_0},\bar p_1,\cdots,\bar p_i,\cdots,\bar p_N,\bar p_{\mathcal{R}}]^T$. In our analysis, we select $\bar{\textbf{p}}$ as the solution of the primary utility maximization problem, i.e; $\bar{p_0}$ is the PU power level that maximizes $U_p$, $\bar p_i$ is the power of the SU$_i$ that maximizes $U_p$ and finally $\bar{p_{\mathcal{R}}}$ is the primary relay power level which maximizes $U_p$.

Based on the above definitions, the primary utility can be written as
\begin{multline} U_p=w_p(1+\gamma_{p_1})(1+\gamma_{p_2})+\sum_{i=1}^N \frac{\gamma_i(\textbf{p})}{2(1+\gamma_i(\textbf{p}))}\\
=w_p(1+\frac{|g_p|^2p_0}{N_{\circ}})(1+\frac{|g_p|^2p_{0_{max}}}{N_{\circ}+|h_0|^2p_\mathcal{R}})\\+ \sum_{i=1}^N \frac{|g_i|^2|h_i|^2p_ip_{\mathcal{R}}}{2(N_{\circ}+|h_i|^2p_\mathcal{R})(N_{\circ} +\sum_{j=0}^N|g_j|^2p_j)}
\end{multline}
and the problem of maximizing the PU utility function can be formulated as follows.
\begin{equation}
\begin{aligned}
& \underset{\textbf{p}}{\text{max}}
& & U_p \\
& \text{subject to}
&& p_i \leq p_{i_{max}}, \; i = 0, \ldots, N\\
&&& \gamma_i \geq \gamma_{i_{th}}, \;  \ i = 1, \ldots, N\\
&&& p_{\mathcal{R}} \leq p_{\mathcal{R}_{max}}.
\end{aligned}
\end{equation}
This problem can be rewritten as
\begin{equation}
\label{geometric}
\begin{aligned}
& \underset{\textbf{p}}{\text{min}}
& & 1/U_p \\
& \text{subject to}
&& p_i/p_{i_{max}} \leq 1, \; i = 0, \ldots, N\\
&&& \gamma_{i_{th}}/ \gamma_i \leq1, \;  \ i = 1, \ldots, N\\
&&& p_{\mathcal{R}}/p_{\mathcal{R}_{max}} \leq1 .
\end{aligned}
\end{equation}
After some simplifications, we can write the objective function as a posynomial over posynomial. We can approximate the posynomial in the denominator into a product of monomials, hence, the problem can be converted in to a geometric program \cite{boyd2004convex}, \cite{boyd2007tutorial}. We will perform this convergence using the iterative algorithm proposed in \cite{singh2012interference}. If the problem is infeasible, the primary relay can ban all SUs from accessing the PU spectrum. In this case, the PU will transmit with a fixed power $p_{0_{max}}$ depending on the assumption made in \cite{singh2012interference} which is not always optimal for the PU utility function to be maximized as will be explained later. To ban SU$_j$ from accessing the spectrum, the relay can simply set its price to $\pi_j \geq \frac{1}{2}$, so the best response of SU$_j$ is to send with a zero power level as has been proved in \cite{ren2011pricing}. 

Next, we propose three different approaches, namely, the optimal, the heuristic and the suboptimal algorithms to maximize the primary user's cost function.

\subsection{The Optimal Scheme:}
\label{Optimal}
\vspace{-1 pt}
Instead of banning all SUs from accessing the spectrum, we can allow a subset of them to access the spectrum. This subset is selected so as to maximize the PU utility function. We should note that it may not be possible to find a subset of SUs to allow their access such that all constraints are satisfied, i.e., empty set case and in this case no SU will access the spectrum. The new optimization problem can be written as
\begin{equation}
\label{optimal problem}
\begin{aligned}
\underset{\textbf{\textit{$s$ $\in$ $\mathcal{S}$}}}{\text{max}}\ \
& \underset{\textbf{p}}{\text{min}}
& & 1/U_p \\
& \text{subject to}
&& p_i/p_{i_{max}} \leq 1, \; i \in \left\{0 \cup s\right\}\\
&&& \gamma_{i_{th}} /\gamma_i \leq1, \;  \ i \in s\\
&&& p_{\mathcal{R}}/p_{\mathcal{R}_{max}} \leq1,
\end{aligned}
\end{equation}
where $\mathcal{S}$ is the set of all subsets of SUs, including the empty set $\left\{{\phi}\right\}$, which means that the PU will access in the absence of any SU transmission.

Each SU has a \rm{QoS} constraint and if it cannot be satisfied the primary relay will ban this SU from accessing the channel. Moreover if the access of the $i$-th SU contradicts with maximizing the primary user utility, the relay will also ban this SU by setting a high price $\pi_i \geq \frac{1}{2}$ for this SU. Hence, this SU$_i$ best response in this case is to not access the channel, i.e, $p_i=0$. 

Optimization over $s$ can be accomplished combinatorially. Each user is represented by a binary value which indicates its state, i.e., active or inactive. Active SU will be indicated by $1$ and inactive SU will be indicated by zero. The possible states are the combination of $N$ binary values with a maximum of $2^N$ possibilities.

 Optimization over \textbf{p} is done using the same technique used in problem (\ref{geometric}). It is clear that the optimal scheme complexity grows exponentially as $N$ increases. The solution of this problem is the desired power allocation vector $\bar{\textbf{p}}$ and then the optimal prices can be calculated through equation (\ref{price}). We denote the maximum value of $U_p$ calculated through problem (\ref{optimal problem}) as $u_p$, which is the maximum utility that can be achieved by any scheme. 
\subsection{The Heuristic Scheme:}
\label{Heuristic}
\vspace{-1 pt}
The optimal scheme, which we have discussed above, becomes more complicated as $N$ increases. Here, we present a simple heuristic scheme which is suitable for real time implementation. In this scheme, the relay chooses only the best SU to access the PU spectrum and bans all other SUs. The best SU is defined as the SU with the maximum harmonic mean ($\mu _H$) \footnote{The subscript $H$ is used throughout this paper to indicate the heuristic scheme.}
of the instantaneous channel gains $|g_i|^2$ and $|h_i|^2$ which can be defined as \cite{springer1979algebra}:
\begin{equation}
\label{mh}
\begin{aligned}
\mu_{H_{i}}=\frac{2|g_i|^2|h_i|^2}{|g_i|^2+|h_i|^2}.
\end{aligned}
\end{equation}
The PU utility when all the SUs are inactive, dented by $u_{P_0}$, can be expressed as follows.
\begin{equation}
\label{Up_0}
\begin{aligned}
u_{p_{0}}=w_p(1+\gamma_{p_{max}})^2.
\end{aligned}
\end{equation}
The PU maximum utility in case that the best SU $j$ is the only SU that accesses the channel ($u_{p_{1}}$) can calculated through the following optimization problem.
\begin{equation}
\label{up1}
\begin{aligned}
& \underset{\textbf{p}}{\text{max}}
& & U_p \\
& \text{subject to}
&& p_0 \leq p_{0_{max}}\\
&&& p_j \leq p_{j_{max}}\\
&&& \gamma_j \geq \gamma_{j_{th}}\\
&&& p_{\mathcal{R}} \leq p_{\mathcal{R}_{max}}.
\end{aligned}
\end{equation}
Hence, we can use Algorithm. \ref{Algorithm 1: Heuristic Scheme} to calculate the maximum PU utility, $(u_{p_{H}})$, for the proposed heuristic approach.
\begin{algorithm}
Calculate $u_{p_{0}}$ using (\ref{Up_0}). \\
Calculate $\mu_{H_{i}}$ for each SU$_i$ using (\ref{mh}). \\
Find the SU with the maximum harmonic mean, $j$.\\
Calculate $u_{p_{1}}$ through (\ref{up1}).\\
Calculate $u_{p_{H}}$ as follows
\begin{equation}
\begin{aligned}
u_{p_{H}}=\max ({u_{p_{1}},u_{p_{o}}}).
\end{aligned}
\end{equation}
\caption{{\sc}}
\label{Algorithm 1: Heuristic Scheme}
\end{algorithm}
Define the relative PU utility achieved by the heuristic scheme with respect to the optimal scheme, which indicates how near is the heuristic scheme from the optimal scheme as follows.
\begin{equation}
\begin{aligned}
r_H= \frac{u_{p_H}}{u_p}.
\end{aligned}
\end{equation}

 \subsection{The Suboptimal Scheme:} 
 \label{Suboptimal}
  \vspace{-1 pt} 
Here, we present a simple suboptimal algorithm. The complexity of this suboptimal scheme is a linear function of $N$, unlike the optimal scheme which has an exponential complexity. Moreover, the performance of the proposed suboptimal scheme lies between that of the optimal and the heuristic schemes as will be shown in Section \ref{Simulation}.

Unlike the heuristic scheme, the suboptimal scheme may allow more than one SU to access the spectrum. The suboptimal scheme can be described as an incremental admission policy, in which the PU gradually adds the SUs one after one according to a certain list provided that adding more SUs will cause an increase in the primary utility.

The problem of maximizing $U_p$ can be reformulated as follows.
\begin{equation}
\begin{aligned}
& \underset{\textbf{p},\rm{A}}{\text{max}}
& & w_p(1+\gamma_{p_1})(1+\gamma_{p_2})+\sum_{i=1}^N \ \frac{\gamma_i(\rm{A}\textbf{p})}{2(1+\gamma_i(\rm{A}\textbf{p}))}\\
& \text{\small subject to}
&& p_i/p_{i_{max}}  \leq 1, \ \; i = 0, \ldots, N\\
&&& a_{i}(\gamma_{i_{th}}/\gamma_i(\textbf{p})) \leq 1, \;  \ i = 1, \ldots, N\\
&&& a_{i} (a_{i}-1) =0 , \;  \ i = 1, \ldots, N\\
&&& p_{\mathcal{R}}/p_{{\mathcal{R}}_{max}} \leq 1,
\end{aligned}
\end{equation}
where ${\textbf{p}}=[{p_0}, p_1, p_2,\cdots, p_N, p_R]^T $ is the power allocation vector, $a_{i}$ is a binary variable and the matrix $\rm{A}$ is a diagonal matrix with $\rm{diag}(\rm{A})$ $=$[$1$, $a_{1},\cdots, a_{i}, \cdots, 1$]. The binary variable $a_{i}$ controls the SU$_i$ QoS constraint; if $a_{i}=1$ this means that SU$_i$ will access the spectrum and its QoS is guaranteed, otherwise it will not access.

Unfortunately, the binary constraint is a non-convex constraint. This problem can be relaxed into the following program which can also be solved using the algorithm proposed in\cite{singh2012interference}:
\begin{equation}
\label{relaxation}
\begin{aligned}
& \underset{\textbf{p},\rm{\hat{A}}}{\text{max}}
& & w_p(1+\gamma_{p_1})(1+\gamma_{p_2})+\sum_{i=1}^N \ \frac{\gamma_i(\rm{\hat{A}}\textbf{p})}{2(1+\gamma_i(\rm{\hat{A}}\textbf{p}))}\\
& \text{\small subject to}
&& p_i/p_{i_{max}} \leq 1 , \ \; i = 0, \ldots, N\\
&&& \hat a_{i}(\gamma_{i_{th}}/\gamma_i(\textbf{p})) \leq 1, \;  \ i = 1, \ldots, N\\
&&& \hat a_{i} \leq 1 \;\ i = 1, \ldots, N\\
&&& p_{\mathcal{R}}/p_{{\mathcal{R}}_{max}} \leq 1,
\end{aligned}
\end{equation}
where $\rm{\hat{A}}$ is the approximate value of $\rm{A}$ calculated after the relaxation (\ref{relaxation}). Define the set $\hat{S}$ as the SUs selected set by the PU to access the spectrum, also define the vector $\hat{\mathbf{a}}$ as follows
\begin{equation*}
 \hat {\mathbf{a}}=[ \hat a_1, \hat a_2, \cdots, \hat a_N].
\end{equation*}
Without loss of generality, we assume that the vector $\hat {\mathbf{a}}$ is sorted in descending order, i.e, $\hat a_1 \geq \hat a_2 \geq \cdots \geq \hat a_N $. After finding the vector $\hat {\mathbf{a}}$ using (\ref{relaxation}), we can use Algorithm. \ref{Algorithlm 2} to find the suboptimal maximum value of the PU utility, $u_{p_{S}}$.
 \footnote{The subscript $S$ is used throughout this paper to indicate the suboptimal scheme.}:
\begin{algorithmic}
\begin{algorithm}
Initialize $\hat{S}_{0}=\left\{\emptyset  \right\}$.\\
Calculate $u_{p_{0}}$ using (\ref{Up_0}). \\
\FOR{k=$1,.., N$}
       \STATE{Solve (\ref{set}) for $S=\left\{\hat{S}_{k-1} \cup k \right\}$ and find $u_{p_k}$} which is the maximum value of the objective function.
       \IF{$u_{p_k} < u_{p_{k-1}}$}
               \STATE $u_{p_{S}}=u_{p_{k-1}}$
               \STATE break;
             \ENDIF
       \STATE {$\hat{S}_{k}=S$}
       \STATE $u_{p_{S}}=u_{p_k}$
\ENDFOR 
\caption{{\sc}}
\label{Algorithlm 2}
\end{algorithm}

\end{algorithmic}

\begin{equation}
\label{set}
\begin{aligned}
& \underset{\textbf{p}}{\text{max}}
& & U_p \\
& \text{subject to}
&& p_i \leq p_{i_{max}}, \; i \in \left\{0 \cup S \right\}\\
&&& \gamma_i(\textbf{p}) \geq \gamma_{i_{th}} , \;  \ i \in  S\\
&&& p_{\mathcal{R}} \leq p_{\mathcal{R}_{max}} .
\end{aligned}
\end{equation}
Similarly, we define the relative utility of the suboptimal scheme with respect to the optimal scheme, which indicates how near is the suboptimal scheme from the optimal scheme, as follows 
\begin{equation}
\begin{aligned}
r_S= \frac{u_{p_S}}{u_p}
\end{aligned}
\end{equation}
\section{Simulation Results}
\label{Simulation}

In this section, we present some numerical simulation results related to the performance
of the proposed schemes. Simulations are done using the GGPLAB simulator\cite{mutapcic2006ggplab}. We assume a secondary network of three users (i.e., $N=3$), $p_{0_{max}}=2$ Watt, $p_{{{\mathcal{R}}}_{max}}=10$ Watt, $p_{i_{max}}=1$ Watt, $N_{\circ}=1$ Watt, $\mathbb{E}[{|g_p|^2}]=0.8$, $\mathbb{E}[{|g_i|^2}]=\mathbb{E}[{|h_i|^2}]=0.8$ $\forall i=1 , \cdots, N$ and $\mathbb{E}[{|g_0|^2}]=\mathbb{E}[{|h_0|^2}]=c \mathbb{E}[{|g_p|^2}] $, where $c$ is a parameter that indicates the relation between the distance between the PT and the relay and the distance between the PT and the PD. The minimum SU QoS requirement $\gamma_{{th}}= -10 \ \rm{dB}$ for all SUs. We investigate the following interference scenarios
\begin{enumerate}
\item Weak Interference Case ($c=0.1$)\\
This means that the distance between the PT and the relay is $3$ times the distance between the PT and the PD.
\item Moderate Interference Case ($c=0.25$) \\
This means that the distance between the PT and the relay is $2$ times the distance between the PT and the PD.
\item Strong Interference Case ($c=1$)\\
This means that the distance between the PT and the relay is the same as the distance between the PT and the PD.
\end{enumerate}
\begin{figure}
  \centering
\includegraphics[width=1\linewidth,height=.3\textheight]{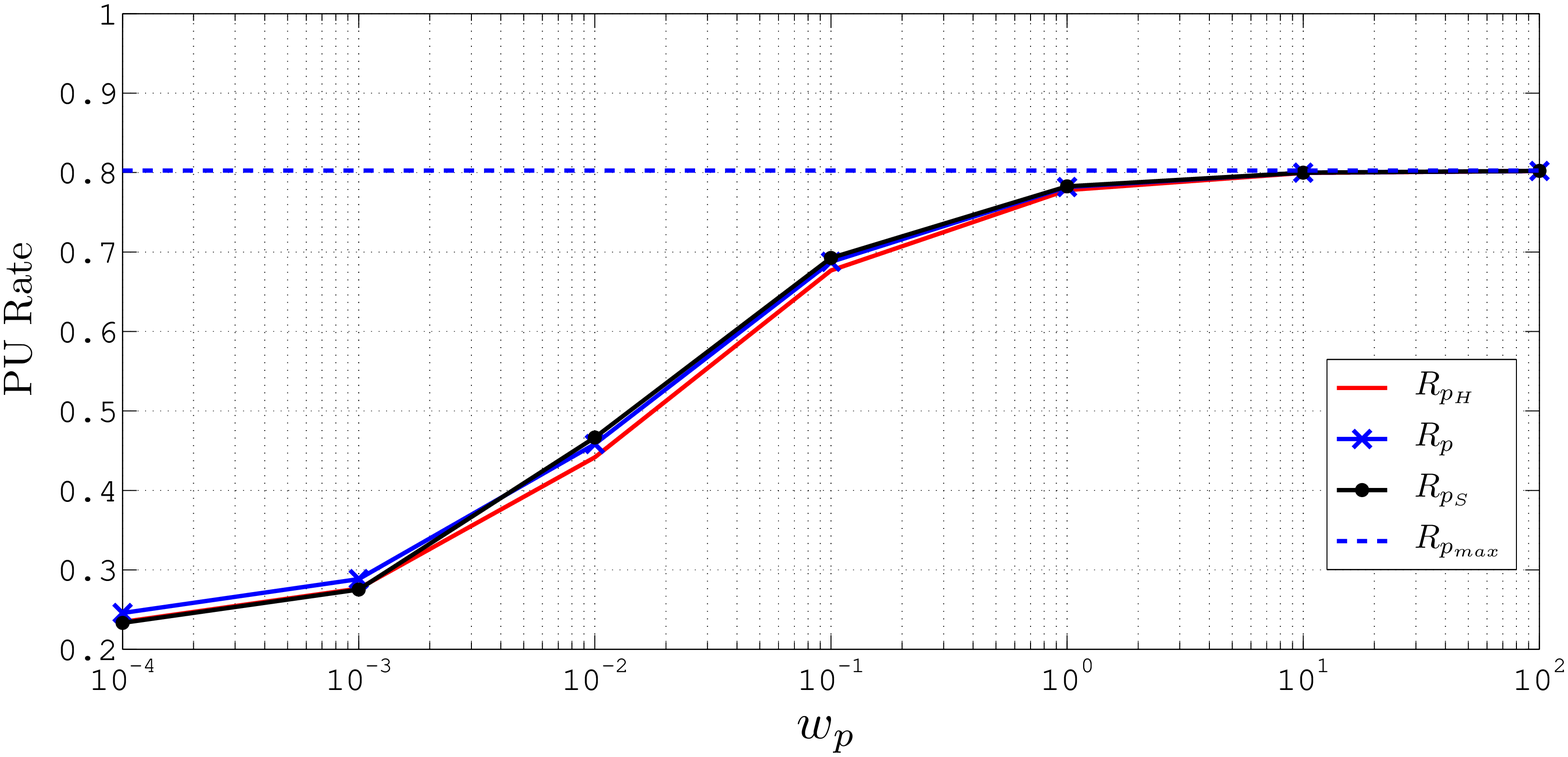} 
\caption{ Primary User Rate as a function of $w_p$, ($c=0.25$) \label{fig 2}}
\end{figure}
\begin{figure}
  \centering
\includegraphics[width=1\linewidth,height=.3\textheight]{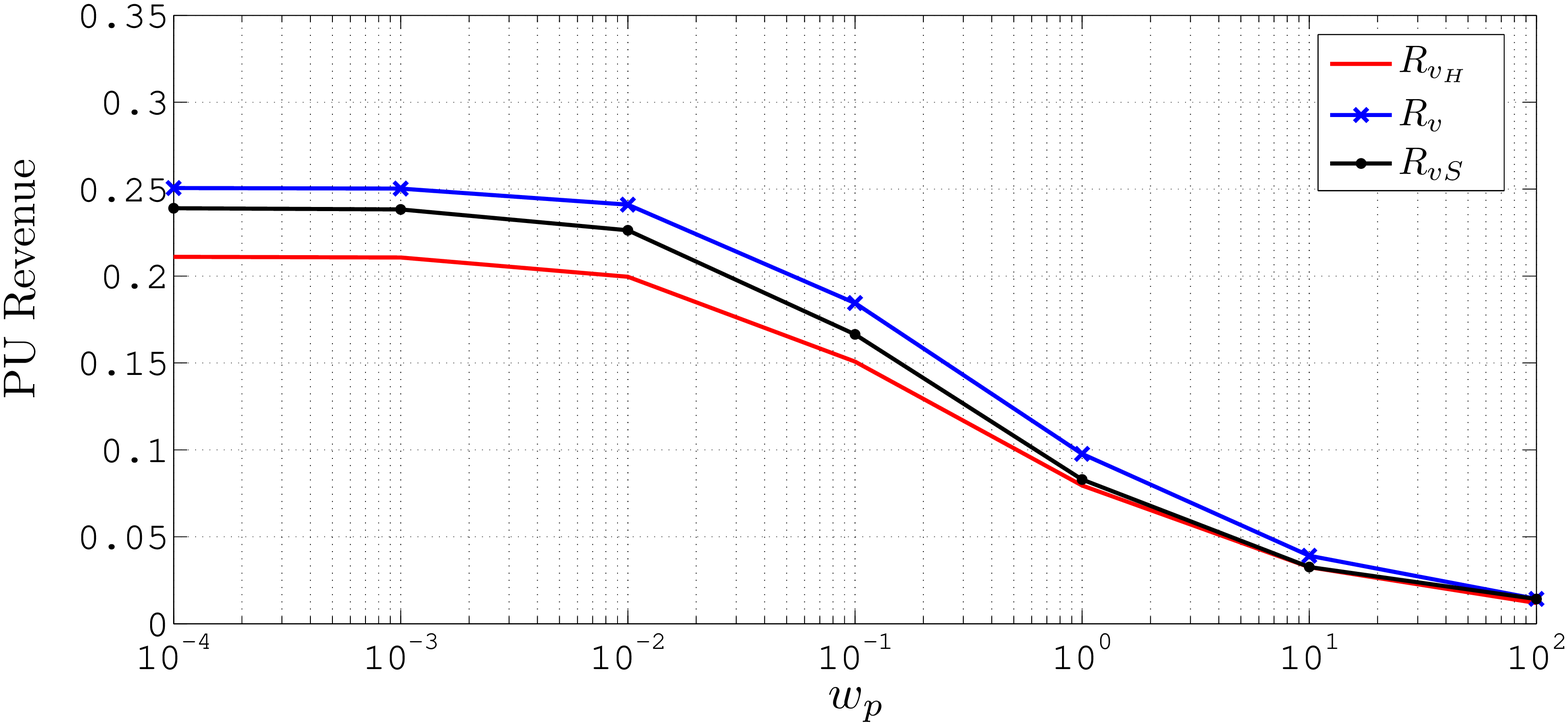} 
\caption{Primary User Revenue as a function of $w_p$, ($c=0.25$)\label{fig 3}}
\end{figure}
\begin{figure}
  \centering
\includegraphics[width=1\linewidth,height=.3\textheight]{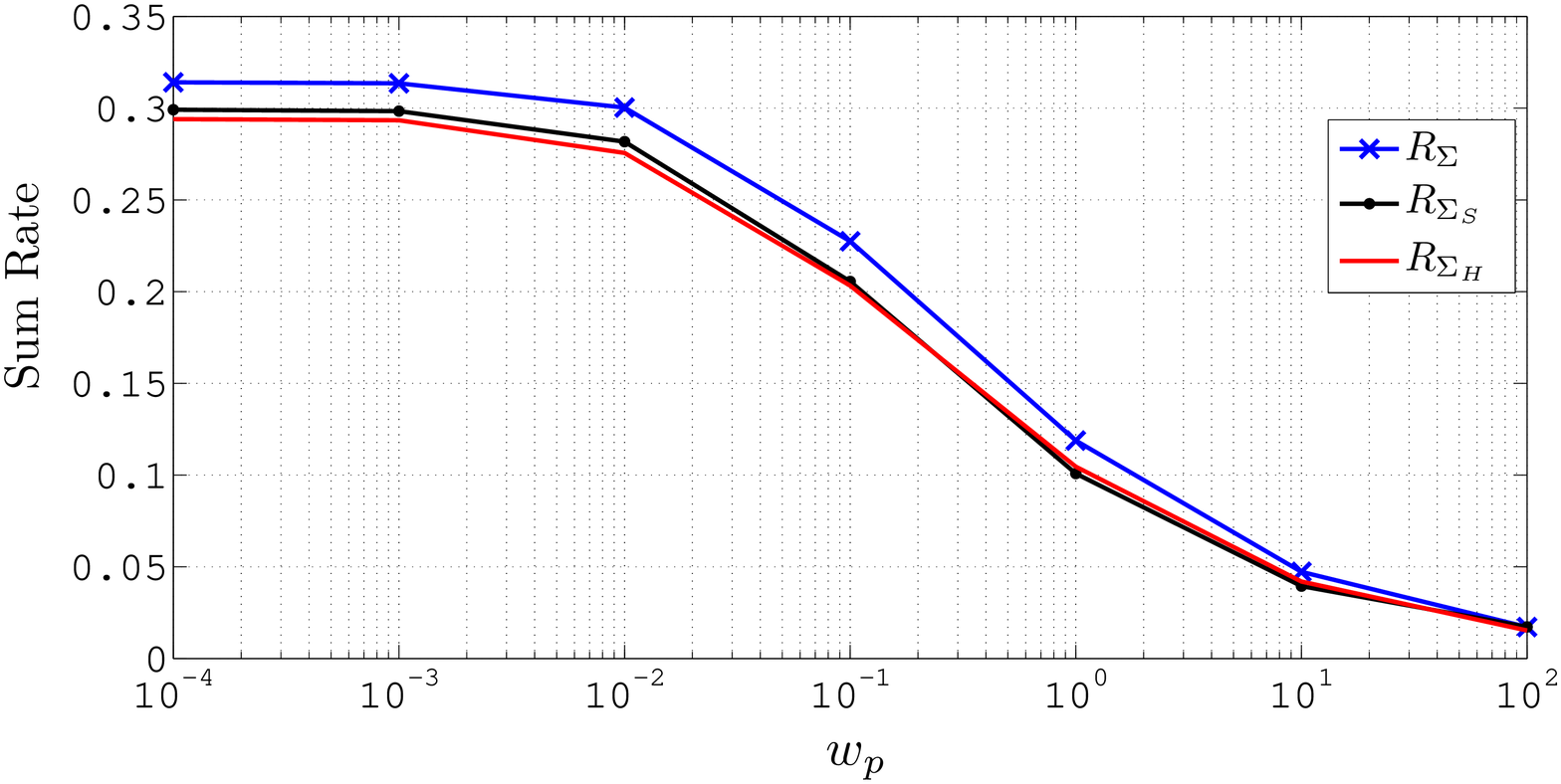} 
\caption{SUs sum rate as a function of $w_p$, ($c=0.25$) \label{fig 4}}
\end{figure}
\begin{figure}
  \centering
\includegraphics[width=1\linewidth,height=.3\textheight]{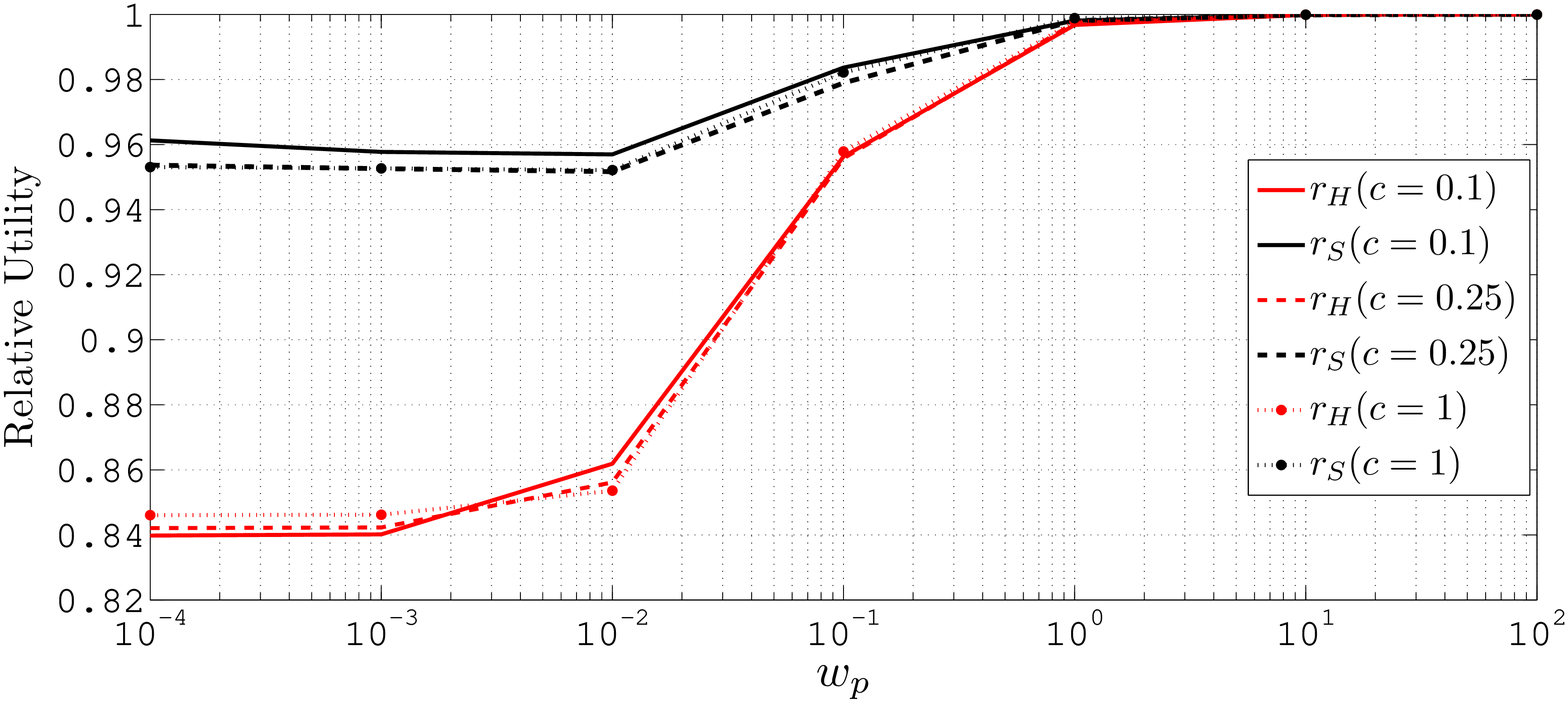} 
\caption{Relative Utility as a function of $w_p$\label{fig 5}}
\end{figure}

\begin{figure}
  \centering
\includegraphics[width=1\linewidth,height=.3\textheight]{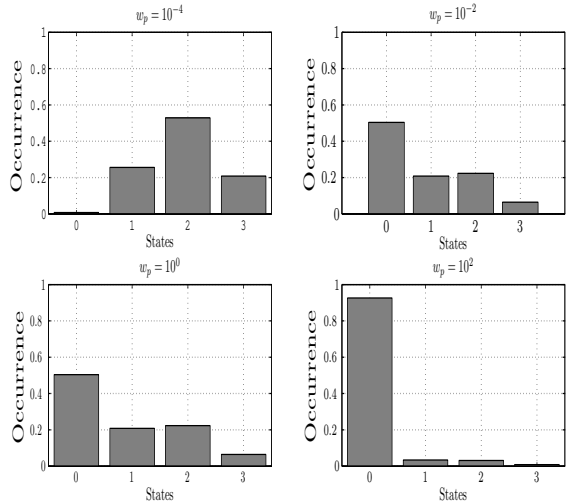} 
\caption{The States Histogram as a function of $w_p$, $(c=0.25)$\label{fig 6}}
\end{figure}
\begin{figure}
  \centering
\includegraphics[width=1\linewidth,height=.3\textheight]{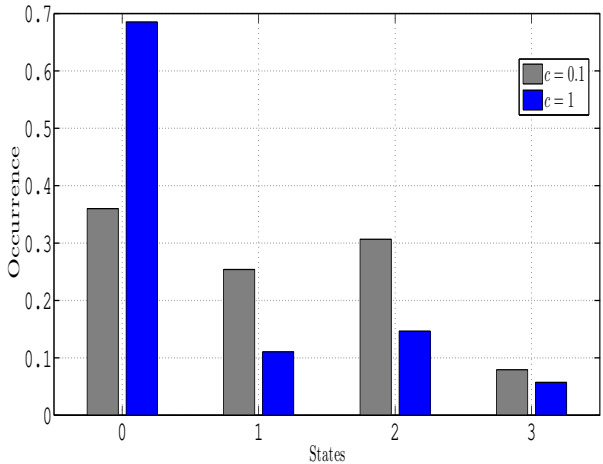} 
\caption{The States Histogram, ($w_p=1$)\label{fig 7}}
\end{figure}
In Fig. \ref{fig 2}, we show that the PU rate is an increasing function of $w_p$. Clearly, as $w_p$ increases, the PU utility function sets more weight to the term $(1+\gamma_{p_1})(1+\gamma_{p_2})$ and hence, the PU rate will increase. It should be noted that as $w_p$ increases the PU rate converges to the maximum PU rate achieved when all SUs are inactive, $R_{p_{\max}}$. In Fig. \ref{fig 3}, we show the primary revenue as a function of the parameter $w_p$, which is a decreasing function of $w_p$. As $w_p$ increases the PU becomes more concerned with its rate rather than its secondary network revenue. It is clear that there is a trade-off between the PU rate and PU revenue and this trade-off can be controlled by the parameter $w_p$. In Fig. \ref{fig 4}, we show that the SU sum rate is decreasing of $w_p$. As $w_p$ increases, the probability of allowing the SUs access decreases since the PU cares more about its achieved rate; this will result in a decrease of the SU sum rate as $w_p$ increases.

In Fig. \ref{fig 5}, the relative utilities, $r_H$ and $r_S$, are shown as function of $w_p$ for different values of the parameter $c$. From that figure, it is clear that the suboptimal scheme outperforms the heuristic scheme because the suboptimal scheme may allow more than one SU to access the spectrum, unlike the heuristic scheme, which allows a maximum of one SU to access the medium. It is also obvious that the closeness of the suboptimal and heuristic schemes to the optimal scheme is almost not affected by the variation of the parameter $c$. 

In Fig. \ref{fig 6}, the occurrence of the different states under the optimal scheme with different values of $w_p$ is shown, where a ``state'' is defined by the number of SUs allowed to access the medium. At low values of $w_p$, the PU permits the access of more SUs since again in this case the PU cares more about it secondary network revenue, whereas at high values of $w_p$, nearly all the SUs are not allowed to access the spectrum most of the time as the PU cares more about its rate. In Fig. \ref{fig 7}, the occurrence of the different states under the optimal scheme for different values of $c$ is shown. It is clear that when the parameter $c$ equals $1$ (Strong Interference) the probability that PU will allow the access of the SUs decreases, as their access in this case significantly affects the PU's transmission. The secondary transmission is also significantly affected by the primary transmission in this case.
\section{Discussion}
\label{Discussion}
In this paper, we have studied a cognitive radio system with one PU, one primary relay and $N$ SUs, in which the PU can control the secondary network through its relay to maximize its desired utility function. Each SU has a QoS constraint and if it cannot be satisfied or it is not beneficial for the PU to allow the access of this SU, this SU will not access the channel. We have investigated the PU trade-off between its achieved QoS and its gained revenue to allow the access of the SUs to its licensed spectrum. We have proposed an optimal power control scheme, which has an exponential complexity in terms of the number of secondary users; therefore, we have also proposed a simple heuristic scheme and a suboptimal scheme which has a linear complexity in terms of the number of the secondary users which achieves a performance that is very close to the optimal scheme. Finally, we have investigated the performance of these schemes relative to the optimal scheme under various interference scenarios, we have concluded that the closeness of these schemes to the optimal scheme is almost the same regardless of the interference scenario.

\bibliographystyle
{IEEEtran}
\bibliography{IEEEabrv,Nulls}

\end{document}